\documentclass[noshowpacs,amsmath,
twocolumn,
superscriptaddress,
8pt,
aps,prb]{revtex4-2}
\usepackage{caption}
\usepackage{subcaption}
\usepackage{lineno}

\usepackage{setspace}
\usepackage{amsmath}
\usepackage{multirow}
\usepackage{graphicx}

\usepackage{verbatim}
\usepackage{amsfonts}
\usepackage{amssymb}
\usepackage{epstopdf}
\usepackage{dcolumn}
\usepackage{xcolor}
\usepackage{verbatim}
\usepackage[colorlinks=true, frenchlinks=false, urlcolor=blue, linkcolor=blue, citecolor=blue]{hyperref}
\usepackage{siunitx}
\usepackage[version=4]{mhchem}
\usepackage{textcomp}
\setcitestyle{super} 
\DeclareGraphicsExtensions{.pdf,.eps,.png,.jpg,.mps}

\begin{document}
\title{Reducing Averaging Time in Dual-comb Spectroscopy via Phase-Patterned Higher-Repetition-Rate Pulses}

\author{Wei Long$^{1,2,*}$, Xinru Cao$^{1,2,3}$, Xiangze Ma$^{1,2,3,4}$, Jiaqi Zhou$^{1,2,3}$, Wenbin He$^{5}$, Dijun Chen$^{1,2,3,4,*}$\\
\medskip
$^1${Wangzhijiang Innovation Center for Laser, Aerospace Laser Technology and System Department, Shanghai Institute of Optics and Fine Mechanics, Chinese Academy of Sciences, Shanghai 201800, China}\\
$^2${Shanghai Key Laboratory of All Solid-State Laser and Applied Techniques, Shanghai Institute of Optics and Fine Mechanics, Chinese Academy of Sciences, Shanghai 201800, China}\\
$^3${Center of Materials Science and Optoelectronics Engineering, University of Chinese Academy of Sciences, Beijing 100049, China}\\
$^4${Hangzhou Institute for Advanced Study, University of Chinese Academy of Sciences, Hangzhou 310024, China}\\
$^5${Russell Centre for Advanced Lightwave Science, Shanghai Institute of Optics and Fine Mechanics and Hangzhou Institute of Optics and Fine Mechanics, Chinese Academy of Sciences, Shanghai 201800, China}\\
$^{*}$longwei20@mails.ucas.ac.cn\\
$^{*}$djchen@siom.ac.cn}

\begin{abstract}

\noindent Dual-comb spectroscopy (DCS) is a powerful Fourier-transform spectroscopic technique that provides high-speed, high-resolution, and broadband measurements without moving parts. However, the high peak power of mode-locked pulses limits the photodetector's dynamic range, resulting in a low signal-to-noise ratio (SNR) per acquisition. While coherent averaging can improve SNR, it sacrifices temporal resolution and demands stringent system stability.  Here, we introduce a novel concept to enhance SNR by using phase-patterned higher-repetition-rate combs. We reinterpret the self-imaging process of comb spectrum from a new perspective on mode interference among sub-pulse trains
As a proof-of-concept, we densified two 250-MHz frequency combs to 12.5-MHz mode spacings via phase modulation and performed DCS on an $\mathrm{H^{13}C^{14}N}$ gas cell,  and compared the results with an emulated conventional 12.5-MHz DCS, demonstrating a 17-fold increase in mode amplitude. This concept is expected to be combined with ultra-high repetition rate combs, such as microcombs, and thereby deployed in practical applications that typically require spectral sampling spacings from hundreds of MHz to GHz range. 

\end{abstract}

\maketitle

\medskip
\section{Introduction}
Enabled by the unique properties of optical frequency combs (OFCs) \cite{20Years2019fortier}, dual-comb spectroscopy (DCS) has emerged over the past decade as a powerful, next-generation Fourier-transform spectroscopy technique, offering an important tool for applications requiring high speed, broad bandwidth, high frequency resolution, and high frequency accuracy, such as in investigations of physical turbulent environments \cite{FrequencycombbasedRemote2014rieker,OpticalPhase2014sinclair,RegionalTracegas2018coburn,SpatiallyResolved2022yuna,DualcombSpectroscopy2024han} and chemical and biological dynamic processes \cite{klocke2018single,DualcombSpectroscopy2020pinkowski,MicrosecondResolvedInfrared2021norahan,UrbanOpenair2023westberga}.
By employing two coherent OFCs with a slight repetition rate difference ($\Delta f_{\mathrm{rep}}$), and inherently free from mechanical moving parts, DCS can capture an interferogram (IGM), yielding a spectrum with a spectral resolution set by the repetition rate of the combs. The time frame for this capture is as short as $1/\Delta f_{\mathrm{rep}}$, and with large $\Delta f_{\mathrm{rep}}$ configurations, this leads to theoretical temporal resolution reaching microsecond or even nanosecond timescales \cite{TimeresolvedMidinfrared2019abbas,11msTimeresolved2021hoghooghi,NanosecondTimeresolved2024long}. 

However, in typical practical DCS systems, the high peak power of mode-locked pulses often leads to photodetector saturation or nonlinear response, thus restricting the average power of the combs to a low level. Consequently, given the vast number of comb teeth inherent to broadband spectra, the power per comb tooth is typically limited to the nanowatt range \cite{SensitivityCoherent2010newbury}, ultimately resulting in extremely low SNR per single acquisition, When the comb source power is sufficient, this is referred to as the dynamic range limitation\cite{DualcombSpectroscopy2016coddington}, which necessitates coherent averaging or multi-period acquisition to enhance the SNR \cite{TimedomainMidinfrared2004keilmann,PhasestableDualcomb2018chen,11msTimeresolved2021hoghooghi}. Furthermore, the SNR gain achieved through averaging scales proportionally to the square root of the number of averages. For instance, to improve the SNR by a factor of 10, a hundred averages are needed. Obviously, this process not only significantly compromises the exceptional time-resolution of DCS by several orders of magnitude, but also places greater demands on the long-term mutual coherence and stability of the system, which greatly increases the implementation complexity and cost.

To mitigate this issue, one approach involves employing sequential spectrum acquisition using multiple photodetectors \cite{SensitivityCoherent2010newbury} or averaging spectral copies instead of IGMs \cite{UnlockingLower2023walsh}. However, the SNR enhancement offered by these two methods has been limited by experimental complexity or sampling bandwidth constraints, demonstrating averaging-time reduction of less than one order of magnitude. More recently, the utilization of  $\sim$ 1-GHz-repetition-rate combs \cite{11msTimeresolved2021hoghooghi,ArchitectureMicrocombbased2021bao,NanosecondTimeresolved2024long,PracticalGHz2025ling} offers higher average power and larger $\Delta f_{\mathrm{rep}}$. Nevertheless, a $\sim$ 1-GHz repetition rate had struck a delicate balance between measurement speed and spectral bin spacing  \cite{Broadband1GHz2022hoghooghi}. A higher repetition rate inherently implies a lower spectral bin spacing, which limits further increases in repetition rate.

Over the past two decades, a comb densification technique based on the spectral self-imaging effect (also known as the spectral Talbot effect) \cite{SpectralTalbot2005azaña} has been developed \cite{SpectralSelfimaging2011caraquitena,ReconfigurableMultiwavelength2011beltran,DiscretelyTunable2013malacarne,ModeSpacingDivision2019li,ArbitraryEnergyPreserving2019romerocortés,UltraDenseCEOStabilized2019li}. This technique applies periodic temporal-phase modulation (TPM) to the comb pulses, thereby inducing evenly spaced "self images" of the original comb modes and leading to the densification of the comb \cite{DualcombSpectroscopy2016hébert,ArbitraryEnergypreserving2018chatellus,DualcombSpectroscopy2024quevedo-galán}. 

In this paper, we model the modulated pulse train as a collection of multiple sub-pulse trains and reinterpret the self-imaging process of comb spectrum from a new perspective on mode interference among these trains. This perspective is analogous to the supermode analysis in harmonic mode-locking\cite{CorrelationSupermode2007gee,OpticalMode2011herra}. Subsequently, we intuitively demonstrate how multiple sub-pulse trains synergistically enhance their common modes. This shows that mode energy can be increased using higher-repetition-rate pulses within the same detector dynamic range without reducing spectral bin spacing, thereby providing an effective solution to the SNR limitation in DCS.

Building on this framework, we provide a complete description for two architectures for this special form of DCS. As a proof-of-concept, we densify two close-to-250 MHz frequency combs via periodic TPM and measured the transmission spectrum of a hydrogen cyanide ($\mathrm{H^{13}C^{14}N}$) gas cell at a spectral sampling bin spacing of 12.5 MHz. Compared to an emulated conventional 12.5-MHz DCS, we demonstrate a 17-fold increase in mode amplitude and an SNR gain of up to $\sim$16.7 times per single acquisition in the non-shot-noise-limited regime, the latter corresponding to a 290-fold reduction in averaging time. This average time reduction is expected to reduce to 17-fold in the shot-noise-limited regime. To distinguish it from conventional dual-comb spectroscopy (C-DCS), we term this approach phase-patterned higher-repetition-rate-pulse enhanced dual-comb spectroscopy (PPHE-DCS).

\medskip
\section{Principles} \label{sec:Chap-principles}

In this section, we first present the spectral behavior of phase-patterned higher-repetition-rate pulses from a perspective of mode interference among sub-pulse trains. Then, two DCS architectures for this PPHE-DCS are studied: one employing a conventional optical frequency comb (C-OFC) and a phase-patterned higher-repetition-rate optical frequency comb (PH-OFC), termed C-PH-DCS; the other employing two PH-OFCs, termed PH-PH-DCS. For clarity, we use a pulse-multiplexing number of \textit{m} = 4 and a TPM sequence of $[\pi,0,0,0]$ as illustrative examples.

\subsection{Spectrum of Phase-Patterned Higher-Repetition-Rate Pulses}

\begin{figure*}[!ht]
    \centering
    \includegraphics[width=1\linewidth]{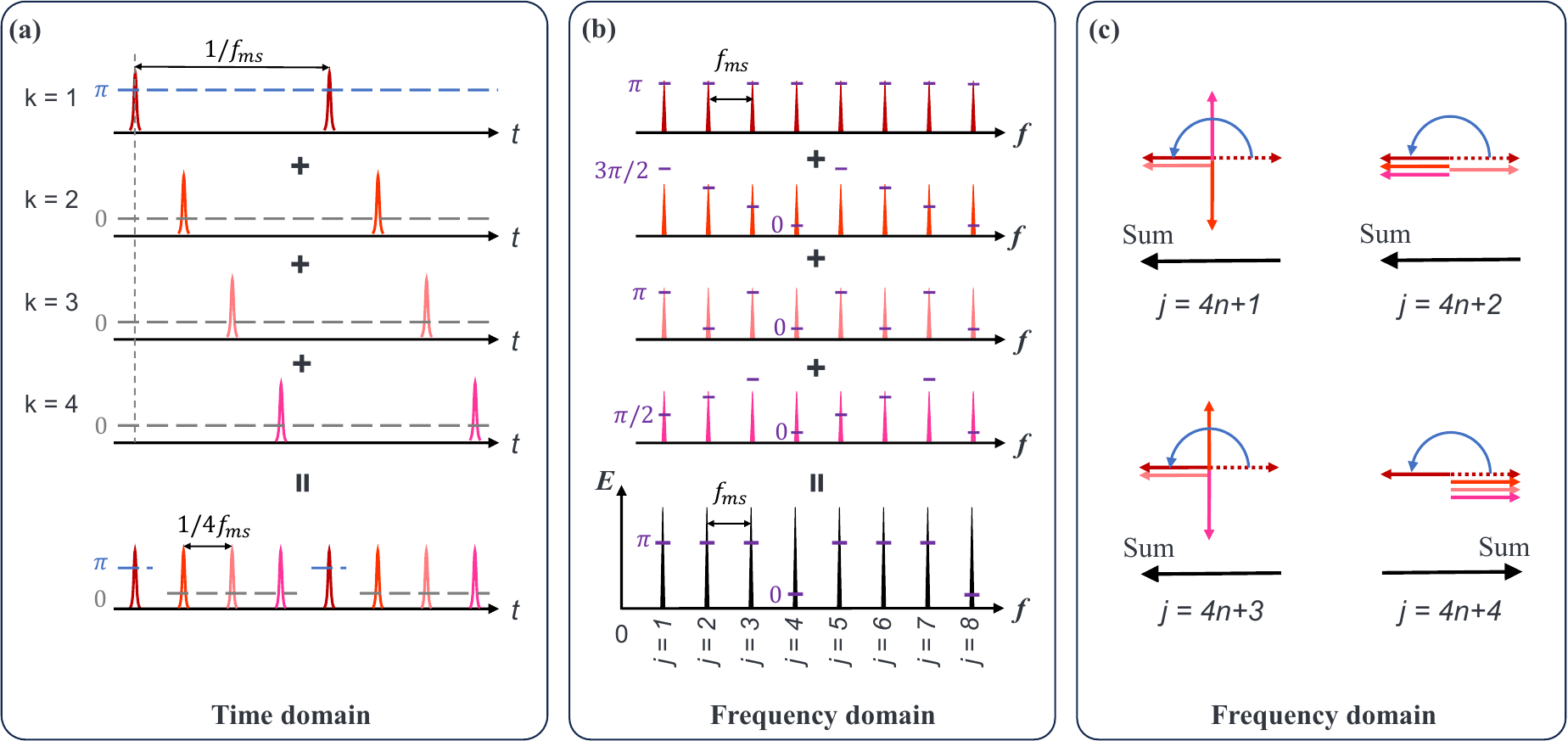}
    \caption{Illustration of the process of the spectral mode superposition in phase-patterned sub-pulse trains. \textbf{(a) }Time domain description. \textbf{(b)} The equivalent frequency-domain description. The \textit{j} represents the spectral mode index (derived from $\omega_j = j \cdot 2\pi f_{\mathrm{ms}}$). The $\pi$ phase of the first spectrum originates from the $\pi$ phase shift in the time domain; the phase slopes of the 2nd to 4th spectra are caused by the phase shift of $-\omega_jt_k$ arising from the temporal shift of pulses. \textbf{(c)} Illustration of vector superposition of the modes of the four sub-pulse trains. The colored linear arrows represent the mode from the \textit{k}th pulse train at $\omega_j$, while the blue rotating arrows indicate the temporal $\pi$-phase shift on the k = 1 pusle train, which rotates all its spectral modes by $\pi$. The black arrows represent the total mode vectors after superposition, which have twice the amplitude of the sub-modes and exhibit a periodic phase profile of $[-\pi, -\pi, -\pi, 0]$. In the absence of the $\pi$-shift, only the modes of $j = 4n$ $(n\in\mathbb{N})$ will be present, while the remaining modes will undergo destructive interference. Consequently, the mode spacing is consistent with the temporal pulse repetition rate, at $4f_{\mathrm{ms}}$, which aligns with classic OFCs. }
\label{fig:sub-pulsetrain-interference}
\end{figure*}

As a example, when phase modulation an OFC with a repetition frequency of $4f_{\mathrm{ms}}$ with a $f_{\mathrm{ms}}$-periodic sequence of $[\pi, 0, 0, 0]$, the original pulse train can be decomposed into a collection of four sub-pulse trains, each with a repetition frequency of $f_{\mathrm{ms}}$, as illustrated in Fig. \ref{fig:sub-pulsetrain-interference}(a). The pulses within each sub-pulse train are still fully coherent, thereby each sub-pulse train corresponding to a sub-spectrum with mode spacing of $f_{\mathrm{ms}}$ (Fig. \ref{fig:sub-pulsetrain-interference}(b)). According to the linear property of the Fourier transform, the spectrum of the PH-OFC or the interferogram at frequency $\omega_j$ is a vector superposition of the 4 sub-modes at that frequency, given by $E_j = \sum_{k=1}^{4}a_0\exp{(\omega _jt+\phi_{k,j})}$, where $a_0$ is the amplitude of the sub-modes. Considering the influence of the relative time delay among these sub-pulse trains and the TPM, the phase spectrum of the \textit{k}th-sub-spectrum can be expressed as
\begin{equation}\label{eq:timeshifting eq}
\phi_{k,j} = -\omega_jt_k+\phi_{TPM,k} = -\frac{j}{4}(k-1)\cdot2\pi+\phi_{TPM,k}.
\end{equation}
The first term in the formula, $-\omega_jt_k$, arises from the contribution of the time shift, where $\omega_j=j\cdot2\pi f_{\mathrm{ms}}$ is the angular frequency of the $j_{th}$ mode, $t_k=(k-1)/4f_{\mathrm{ms}}$ is the time delay of \textit{k}th-sub-pulse train relative to the first sub-pulse train; the second term, $\phi_{TPM,k}$, is the phase shift added to the \textit{k}th-train through the TPM. The superposition process and results are illustrated in Figs. \ref{fig:sub-pulsetrain-interference}(b) and (c), with vector sum $E_j = 2a_0\exp{(\omega _jt+\varphi_{j})}$ where $\varphi_{j}$ is a periodic sequence $[\pi,\pi,\pi,0]$. This indicates that the spectrum of the phase-patterned higher-repetition-rate pulses train has the same mode spacing but 2-fold the amplitude of the sub-spectrum (which directly corresponds to a 4-fold increase in pulses, i.e., a 4-fold increase in optical power), and exhibits a periodic multilevel phase spectrum of $[\pi,\pi,\pi,0]$. This form of spectrum can be viewed as the result of interleaving four independent modesets, within which the mode spacing is $f_{\mathrm{ms}}$ and the modes are perfectly phase-coherent.

In practice, a deviation in the phase modulation level leads to inconsistent interference of modes at different frequencies when all the sub-spectra are superimposed (imaging that the rotation angle of the blue arrow in Fig. \ref{fig:sub-pulsetrain-interference}(c) is not exactly 180 degrees), which in turn causes periodic mode-to-mode-amplitude variations. This also corresponds to the more general case of using an arbitary signal and the resulting spectrum can still be viewed as the result of interleaving four independent modesets.

When the period of the modulation signal is greater than 4, to obtain a consistent constructive interference, one can use either the quadratic function form indicated by the so-called "Talbot condition" \cite{GeneralityTalbot2016cortés} or a binary pseudo- random signal\cite{ModeSpacingDivision2019li}. The  equation of the former is expressed as
\begin{equation}\label{eq:Tbt condition1}
\phi_{n;s,m} = \pi\frac{s}{m}k^2 (mod ~2\pi), 
\end{equation}
where \textit{s}, \textit{m} are mutually prime integer numbers, and k is the pulse number. Correspondingly, the periodic multilevel phase profile of the total spectrum is $\varphi_{k;p,q} = -\pi\frac{p}{q}j^2 (mod ~2\pi),$ where $q = m$, \textit{j} is the mode number, and $p\ne s$ in general. 

\begin{figure*}[!ht]
\captionsetup[subfigure]{labelformat=empty}
    \centering
    \begin{subfigure}{1\textwidth}
      \includegraphics[width=1\textwidth]{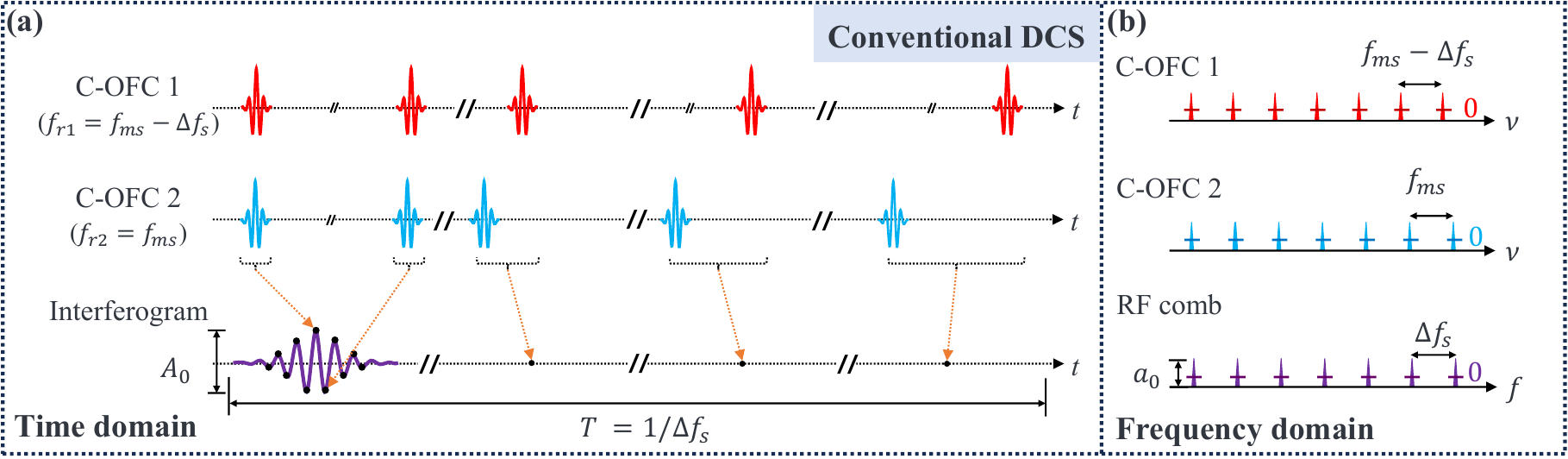}
    \end{subfigure}
    \\
    \begin{subfigure}{1\textwidth}
      \includegraphics[width=1\textwidth]{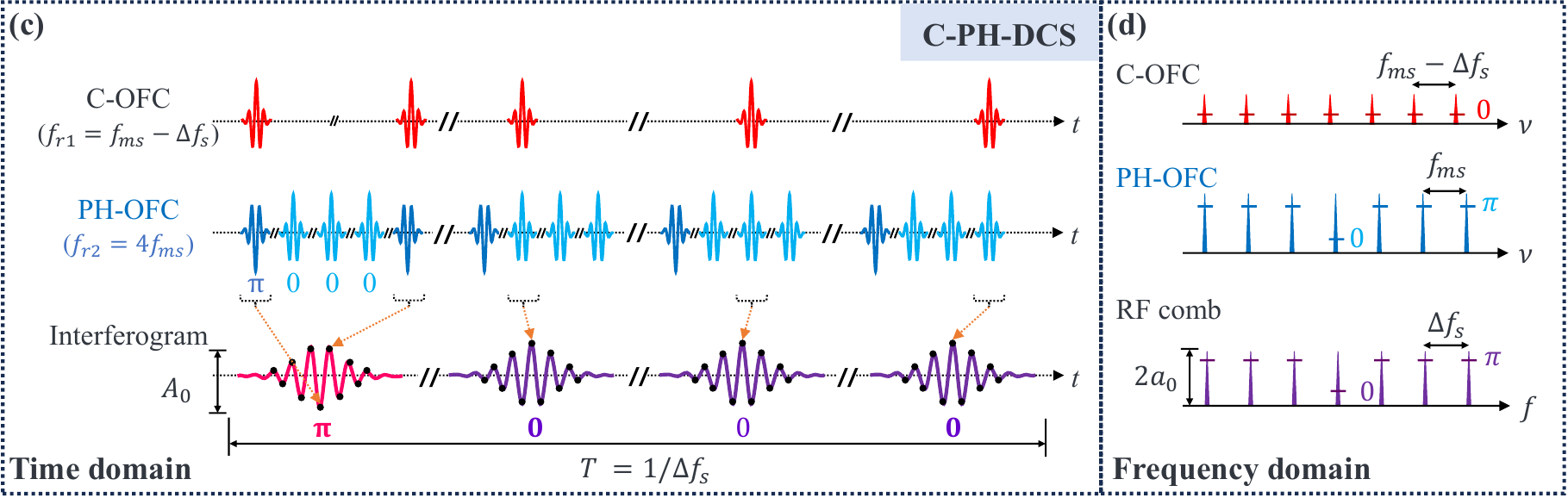}
    \end{subfigure}
\caption{Concepts of conventional DCS (C-DCS) and C-PH-DCS. 
{\bf (a)} Time-domain description of C-DCS. The pulse train (red) of comb 1 progressively steps across another pulse train (blue) with a slight repetition rate difference $\Delta f_{\mathrm{s}}$, generating an interferogram  with a single centerburst in a period of $1/\Delta f_{\mathrm{s}}$. When the temporal interval between the two pulses is large, effective interference does not occur. 
{\bf (b)} The equivalent frequency-domain description of the C-DCS. The beating of two combs produces an RF comb in the radio-frequency domain. 
{\bf (c)} Time-domain description of the C-PH-DCS. The pulse of the conventional optical frequency comb (C-OFC) with a pulse repetition rate of $f_{\mathrm{ms}}-\Delta f_{\mathrm{s}}$, steps across the four pulses of the phase-patterned higher-repetition-rate optical frequency comb (PH-OFC), which have phases of $[\pi, 0, 0, 0]$ and a pulse repetition rate of $4f_{\mathrm{ms}}$. This generates four centerbursts within an interferogram period of $1/\Delta f_{\mathrm{s}}$, which inherit the $[\pi, 0, 0, 0]$ phases. 
{\bf (d)} The equivalent frequency-domain description of the C-PH-DCS. Two combs with mode spacing of $f_{\mathrm{ms}}-\Delta f_{\mathrm{s}}$ and $f_{\mathrm{ms}}$ respectively beat together, producing an RF comb in the radio-frequency domain. This RF comb inherits the $[\pi, \pi, \pi, 0]$ periodic multilevel phase profile of the PH-OFC and exhibits twice the line amplitude compared to the C-DCS.}
\label{fig:C-PH-DCS}
\end{figure*}

\begin{figure*}[!ht]
    \centering
    \includegraphics[width=1\textwidth]{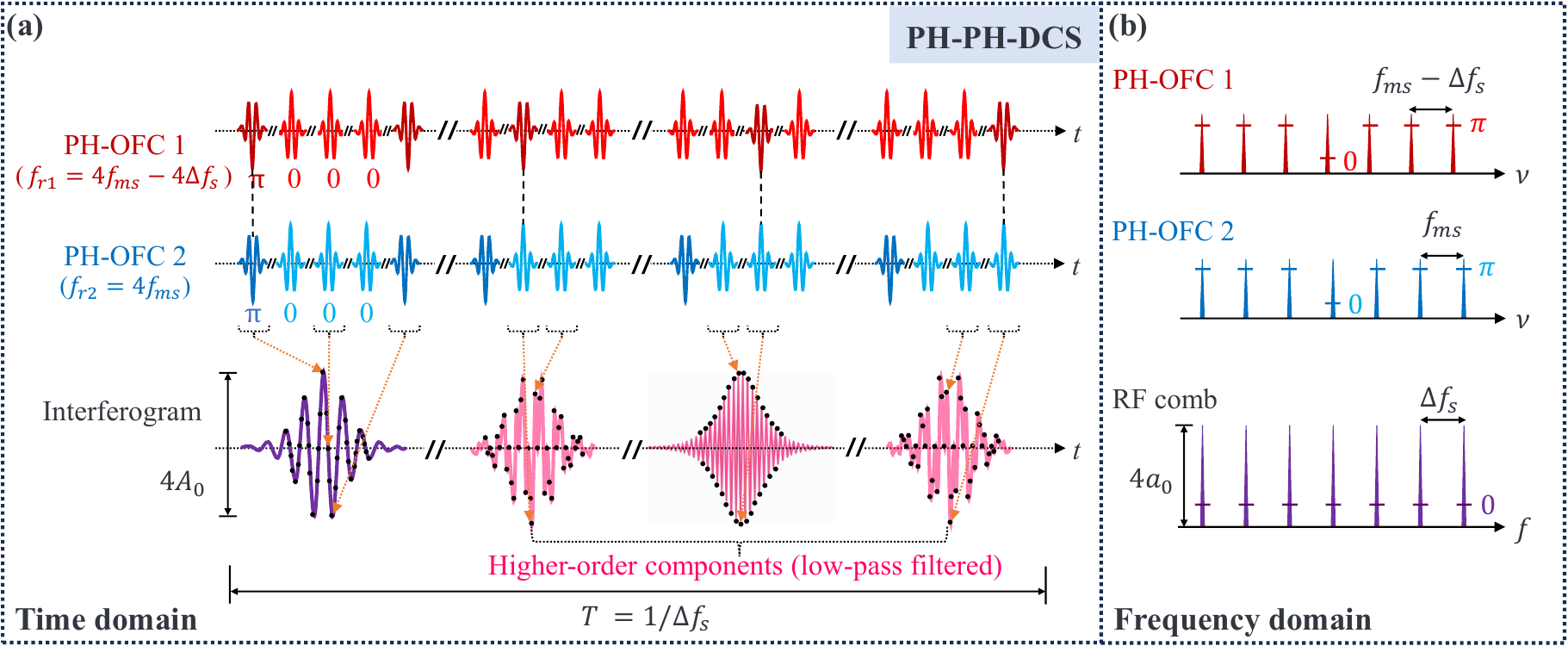}
    \caption{Concept of PH-PH-DCS. {\bf (a)} Time-domain description of the PH-PH-DCS. The PH-OFC 1 and PH-OFC 2 emit two pulse trains with repetition rate of $4(f_{\mathrm{ms}}-\Delta f_{\mathrm{s}})$ and $4f_{\mathrm{ms}}$ respectively, both with pulse phases of $[\pi, 0, 0, 0]$. Interference of the two pulse trains within a time period of $1/\Delta f_{\mathrm{s}}$ produces four distinct centerbursts: one at the fundamental oscillation frequency (purple), exhibiting a 4-fold amplitude enhancement over conventional dual-comb spectroscopy (C-DCS); and three additional centerbursts at higher-order frequencies (pink), which are suppressed by low-pass filtering in practical implementations. {\bf (b)} The equivalent frequency-domain description of the PH-PH-DCS. Two PH-OFCs with mode spacing of $f_{\mathrm{ms}}-\Delta f_{\mathrm{s}}$ and $f_{\mathrm{ms}}$ respectively beat together, producing an RF comb in the radio-frequency domain which has a uniform phase profile and exhibits four times the line amplitude compared to the C-DCS. Note that, the uniform phase profile requires the phase profiles of the two PH-OFCs to be accurately aligned.}
    \label{fig:PH-PH-DCS}
\end{figure*}

\begin{figure*}[!ht]
    \centering
    \includegraphics[width=1\linewidth]{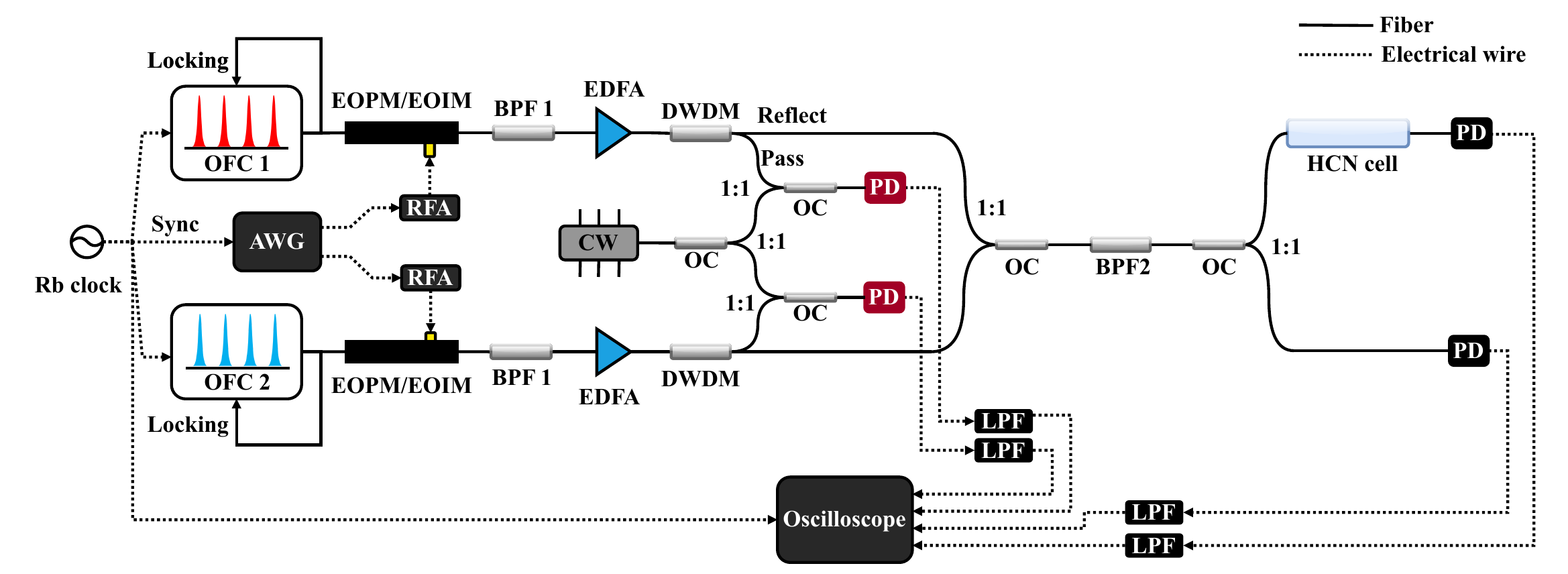}
    \caption{Experimental setup. Rb clock: rubidium clock; OFC: optical frequency comb; AWG: arbitrary waveform generator; RFA: RF amplifiers; EOPM/EOIM: electro-optical phase/intensity modulator; OC: optical coupler; PD: photodetector; LPF: low-pass filter; BPF: band-pass filter; EDFA: erbium-doped fiber amplifier; CW: continuous wave laser; DWDM: dense wavelength-division multiplexer.}
    \label{fig:Experimental setup}
\end{figure*}

\subsection{C-PH-DCS}

As illustrated in Figs. \ref{fig:C-PH-DCS}(c) and (d), C-PH-DCS utilizes a PH-OFC whose repetition rate is approximately equal to 4 times of the repetition rate of the C-OFC (or, more generally, any integer multiple \textit{m}). The pulses of comb 1, after stepping across the first pulse of the PH-OFC, subsequently scan its 2nd to 4th pulses. Consequently, four centerbursts are generated within a single period of $1/\Delta f_{\mathrm{s}}$, as opposed to the single one generated in C-DCS (Fig. \ref{fig:C-PH-DCS}(a)). Additionally, according to the principle of linear optical sampling \cite{SpectrometryFrequency2002schiller,CoherentLinear2009coddingtona}, the expression for the interference intensity can be written as 
\begin{equation}\label{eq:interference eq}
I(n) \sim\sum_{n}^{} \langle | E_1(t-n \tau)| | E_2(t)\left|\cos \left(\Delta \omega_ct+ \phi_{TPM}\right)\right\rangle,
\end{equation}
where $E_1(t), E_2(t)$ represents the pulses' electric field envelopes; \textit{n} denotes the pulse pair number; $\tau$ denotes the time delay induced by the repetition rate difference; and $\phi_{TPM}$ represents the phase difference introduced by the TPM. For each centerburst, the $\phi_{TPM}$ is constant. Consequently, the TPM imposed on the pulses of PH-OFC is finally mapped onto the four centerbursts. 

The spectral behavior of the interferogram with four phase-patterned centerbursts can be also deduced from the spectral mode superposition process as illustrated in Fig. \ref{fig:sub-pulsetrain-interference}. Consequently, the C-PH-DCS (Fig. \ref{fig:C-PH-DCS}(d)) provides a 2-fold amplitude-enhancement in spectral mode compared to the C-DCS (Fig. \ref{fig:C-PH-DCS}(b)), all while preserving the spectral sampling spacing. For a more general case, the number 2 aforementioned corresponds to $\sqrt m$, where \textit{m} is defined as the multiplexing factor of the pulses.

\subsection{PH-PH-DCS}

As illustrated in Fig. \ref{fig:PH-PH-DCS}, PH-PH-DCS utilizes two PH-OFCs with pulse repetition rates of $4f_{\mathrm{ms}}$ and $4(f_{\mathrm{ms}}-\Delta f_{\mathrm{s}})$. In the time domain, similar to a C-DCS with the same repetition rate difference of $4\Delta f_{\mathrm{s}}$, this architecture generates 4 centerbursts within a time period of $1/\Delta f_{\mathrm{s}}$. Due to this repetition rate difference, the phase-inverted pulses in the two pulse trains gradually slip relative to each other, resulting in distinct in-pulse-pair phase configurations: $[\pi, 0, 0, 0] - [\pi, 0, 0, 0]$, $[0, \pi, 0, 0] - [\pi, 0, 0, 0]$, $[0, 0, \pi, 0] - [\pi, 0, 0, 0]$, and $[0, 0, 0, \pi] - [\pi, 0, 0, 0]$, respectively, in the time domain regions corresponding to the four centerbursts. Consequently, the $\phi_{TPM}$ in Eq. \ref{eq:interference eq}, is constantly 0 for the first centerburst; while for the other three centerbursts, the values are periodic $[\pi, \pi, 0, 0], [\pi, 0, \pi, 0]$, and $[\pi, 0, 0, \pi]$, respectively. These time-varying phases (related to the pulse pair index, i.e., time) increase the oscillation frequency of the interference electric field, shifting the 2nd to 4th centerbursts to higher-order components, which can be filtered out by an electronic low-pass filter in practical implementation. In fact, these high-order components correspond to spectral copies generated by the heterodyning of teeth from comb 1 with non-nearest teeth from comb 2 in the spectrum, as detailed in Supplement 1.

Thus, the IGM exhibits only one fundamental centerburst within a time period of $1/\Delta f_{\mathrm{s}}$, similar to that in the C-DCS. Distinctly, the fundamental centerburst originates from the contribution of 4 times the number of pulse pairs, whose interference intensity is given by $I'(n) \sim\sum_{n=1}^{4n_0} \langle | E_1(t-n \tau/4)| | E_2(t)|\cos \Delta \omega_ct\rangle$, where $n_0$ is the number of pulse pairs contributing to the centerburst in the C-DCS illustrated in Fig. \ref{fig:C-PH-DCS}(a). Consequently, following convolution with the detector response and low-pass filtering, the amplitude of this centerburst is also enhanced by a factor of 4.

In the frequency domain, two PH-OFCs with 2-fold mode amplitude enhancement compared to the C-OFC and periodic multilevel phase spectra of $[\pi,\pi,\pi,0]$, beat together to yield an RF comb. Since the amplitude of a heterodyne beat note is the product of the amplitudes of the two beating components, and the phase is the phase difference between them, the modes of this RF comb therefore exhibit a 4-fold enhancement in amplitude and possess a uniform phase, which precisely corresponds to the single fundamental centerburst with a 4-fold enhancement in amplitude in the time domain. For a more general case, the amplitude enhancement is \textit{m}-fold.

A potential misconception might arise that the amplified centerburst amplitude necessitates a larger detector dynamic range. In fact, this is a superposition resulting from the convolution of the additional pulse pairs due to low-pass filtering.  
Interestingly, the aforementioned IGM characteristics can be changed when the two TPM signals use the quadratic form described in Eq. \ref{eq:Tbt condition1} with the same \textit{m} but different \textit{s} values. The generated RF comb can exhibit quadratic periodic multilevel phase profiles of $-\pi\frac{p_1-p_2}{m}j^2$. According to the temporal self-imaging effect \cite{maramNoiselessIntensityAmplification2014,ArbitraryEnergyPreserving2019romerocortés}, the fundamental centerburst on the IGM will then be redistributed to \textit{m}/2 copies, similar to that in C-PH-DCS. A similar effect is expected when using two PRBS signals with different patterns. This will help to alleviate the dynamic range pressure on components within the detection chain other than the detector.

\section{Experiment} 

The scheme of a proof-of-concept experiment with the three architectures is illustrated in Fig. \ref{fig:Experimental setup}. This experiment is conducted using two commercially available 250-MHz fiber mode-locked combs at 1550 nm (Menlo Systems, 250-ULN and 250-WG-PM). The TPM periods were set to 20 times the pulse period in the C-PH-DCS and PH-PH-DCS experiments, yielding a pulse-multiplexing number of 20 and a spectral bin spacing of 12.5 MHz. The TPM is executed by an EOPM with 10-GHz bandwidth. Given a lack of high-quality C-OFCs operating at a repetition rate of 12.5 MHz, and a need for control over variables such as pulse energy and power, the two 12.5 MHz C-OFCs are emulated by implementing the pulse-picking technique \cite{TunableResolution2016vieira, LowrepetitionrateOptical2024canella} with a picking factor of 1/20 applied to the original pulse trains, through electro-optic intensity modulators (EOIMs) with a bandwidth of 10 GHz. An arbitrary waveform generator (AWG, Keysight-M8190A) with a bandwidth of 4 GHz and operating at a sampling rate of 10 GHz was emloyed to generate modulation signals for these modulators. Specifically, the EOIMs were driven by periodic square-wave signals with a 3$\%$ duty cycle (below 5$\%$ to ensure better suppression of adjacent pulses) and repetition frequency of 12.5 MHz, while the EOPMs were driven by the quadratic form referenced in Eq. \ref{eq:Tbt condition1}, using \textit{m} = 20 and \textit{s} = 1 as parameters. Two RF amplifiers (Mini-Circuits ZHL-10M4G21W0+) amplified the modulation signals to an appropriate voltage level required for modulator operation. 

The two OFCs operated at repetition rates of 250 MHz and 249.98 MHz with a $\Delta f_{\mathrm{rep}}$ = 20 kHz, and carrier-envelope offset frequencies ($f_{\mathrm{ceo}}$) of 20 MHz and 33.5 MHz, respectively. To ensure long-term stability of the system, these $f_{\mathrm{ceo}}$ and $f_{\mathrm{rep}}$ were simply phase-locked to a rubidium atomic clock, with which the AWG was also timing synchronized. Variable delay lines or RF phase shifters can be employed to align the optical pulses and modulation signals. The high degree of mutual coherence essential for DCS was retrieved through the optical referencing and post-treatment techniques\cite{ModeresolvedDualcomb2021yu}. A continuous wave (CW) laser is used as an intermediate oscillator \cite{OpticalReferencing2010deschênes} and the self-correction algorithm based on cross-correlation function \cite{SelfcorrectedChipbased2017hébert,SelfCorrectionLimits2019hebert} are implemented.

Furthermore, operating at spectral bin spacing of 12.5 MHz leads to a limited available spectral bandwidth. According to the DCS non-aliasing condition \cite{SensitivityCoherent2010newbury}, the available bandwidth satisfies $\Delta \nu \le f_\mathrm{ms}^2/{2\Delta f_\mathrm{s}}$, which yields 78.125 GHz (0.626 nm at 1550 nm) here for $f_\mathrm{ms}$ = 12.5 MHz and $\Delta f_\mathrm{s}$ = 1 kHz. Therefore, optical band-pass filters (BPFs) were employed to narrow the spectrum, specifically two 1.3-nm @ 25 dB filters (BPF 1) and a 0.3-nm @ 12 dB filter (BPF 2). Because most of the optical power was lost during filtering, two EDFAs were used to partially compensate. The optical power incident on the detector in the measurement beam was $11.9~\mu W$, $119 ~\mu W$, and $223 ~\mu W$ for the C-DCS, the C-PH-DCS, and the PH-PH-DCS.

For the DCS measurement, an $\mathrm{H^{13}C^{14}N}$ gas cell (Wavelength References, HCN-13-10) with a nominal pressure of approximately 10 Torr and a path length of 16.5 cm was selected as a representative sample. In addition to the measurement beam, a reference beam was directly detected for baseline normalization.  Both beams were detected by two InGaAs biased photodetectors (Thorlabs DET01CFC, 1 GHz bandwidth, no internal amplification). The photocurrent was low-pass filtered (6 MHz @ 3 dB)  and then converted to a voltage signal by terminating the detector directly into the 50 $\Omega$ input of a Tektronix MSO54 oscilloscope. The signal was then digitized at a sampling rate of 31.25 MS/s with 12-bit resolution. 

\section{Results and discussion}

\begin{figure*}[!htb]
    \centering
    \includegraphics[width=1\textwidth]{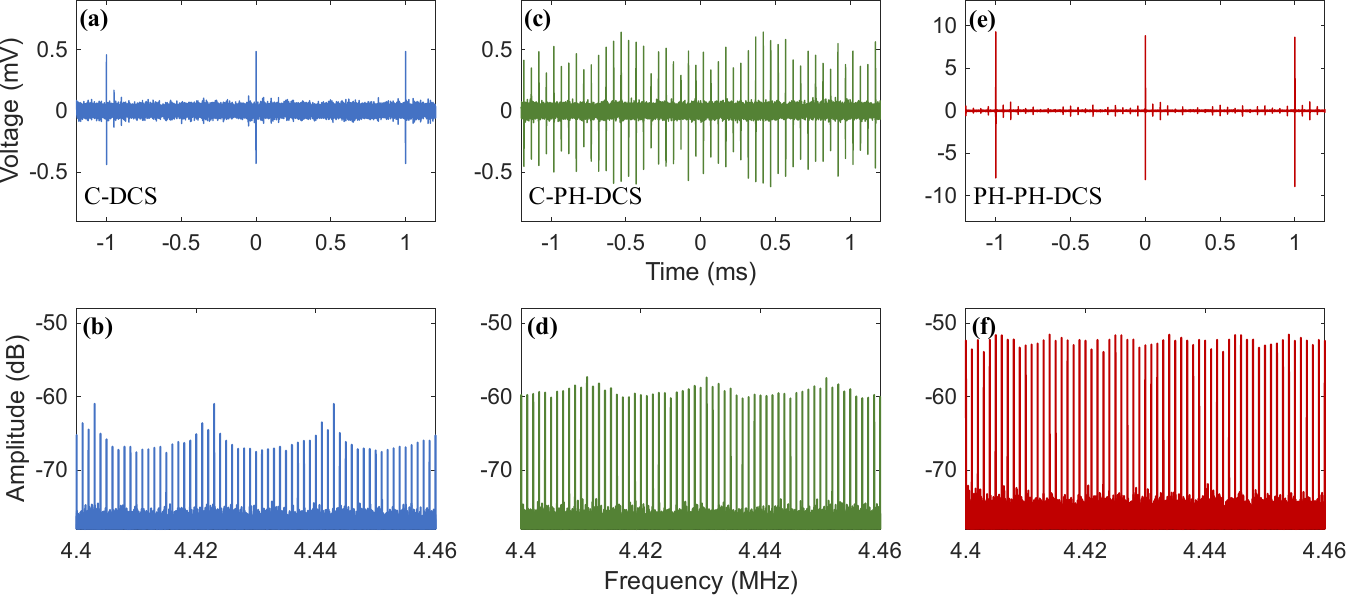}
    \caption{Comparison of interferograms (IGMs) and spectrum for the three architectures. {\bf (a)} IGMs of the C-DCS and {\bf (b)} its spectrum (partial). The centerburst repetition rate is 1 kHz, matching the spectral line spacing, as expected for conventional DCS. The spectrum exhibits a 20 kHz amplitude periodicity, with prominent lines attributed to the non-ideal extinction ratio of the EOIMs \cite{TunableResolution2016vieira}. {\bf (c)} IGMs of the C-PH-DCS and {\bf (d)} its spectrum. The centerburst repetition rate is observed at 20 kHz, a 20th harmonic of its line spacing of 1 kHz, corresponding to the average mode amplitude in the spectrum increasing to $>$-60 dB. {\bf (e)} IGMs of the PH-PH-DCS and {\bf (f)} its spectrum. The centerburst repetition rate and the line spacing are both observed to be 1 kHz. The amplitude of the centerbursts is significantly increased to a value approaching 20 mVpp, corresponding to the spectral mode amplitude increased to close to -50 dB.}
    \label{fig:IGMs&Spec of 3 archis}   
\end{figure*}

\subsection{Features of Interferograms}

Figure \ref{fig:IGMs&Spec of 3 archis}(a–f) displays the raw IGMs generated in the three architectures and their corresponding spectra after post treatment. As anticipated, the centerburst repetition rate in the C-PH-DCS is 20 times higher than that of the C-DCS, while maintaining the same spectral mode spacing and achieving an average enhancement of the mode amplitude by approximately 7 dB ($\mathrm{10lg\sqrt{20}}$).

As for the PH-PH-DCS, the centerburst amplitude is approximately 20 times greater than that in the C-DCS. This corresponds to an average enhancement of the mode amplitude by approximately 13 dB ($\mathrm{10lg20}$) in the spectrum. The noise floor in the spectrum also increases by about 2 dB. Furthermore, in addition to the primary centerbursts, the IGMs of the PH-PH-DCS also show weak centerbursts with a 1/20 ms interval. This is attributed to an error in the modulation signal level, which prevented the original centerbursts at these adjacent locations from being completely transferred to high-frequency components (Fig. \ref{fig:PH-PH-DCS}(a)), leaving a small residue at the fundamental oscillation frequency. The mode superposition of these weak centerbursts caused the unflatness on the RF comb, i.e., an amplitude modulation with a period of 20 kHz, as shown in Figs. \ref{fig:IGMs&Spec of 3 archis}(e) and (f). Details regarding the higher-order components are discussed in Supplement 1. 

\subsection{Signal-to-Noise Ratio and Noise Floor}

\begin{figure*}[!ht]
    \centering

    \begin{subfigure}{0.4\textwidth}
        \includegraphics[width=1\linewidth]{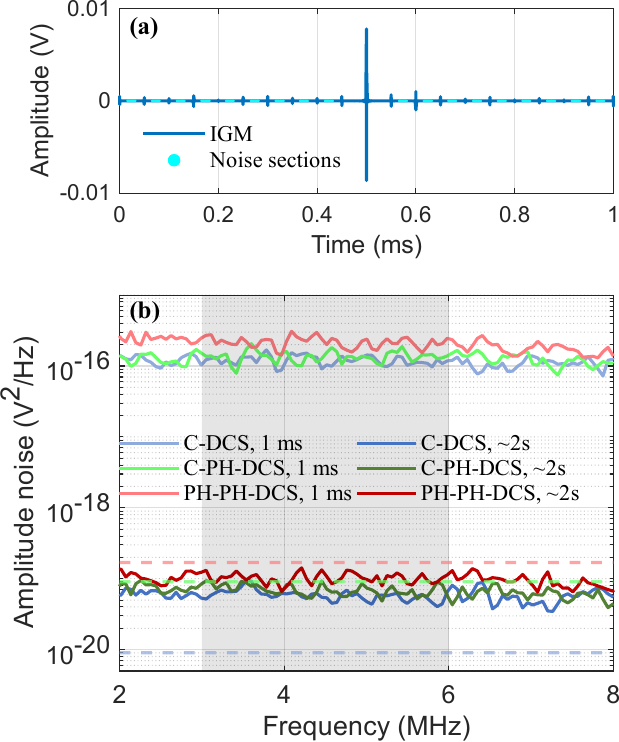}
    \end{subfigure}
     ~
    \begin{subfigure}{0.4\textwidth}
        \includegraphics[width=1\textwidth]{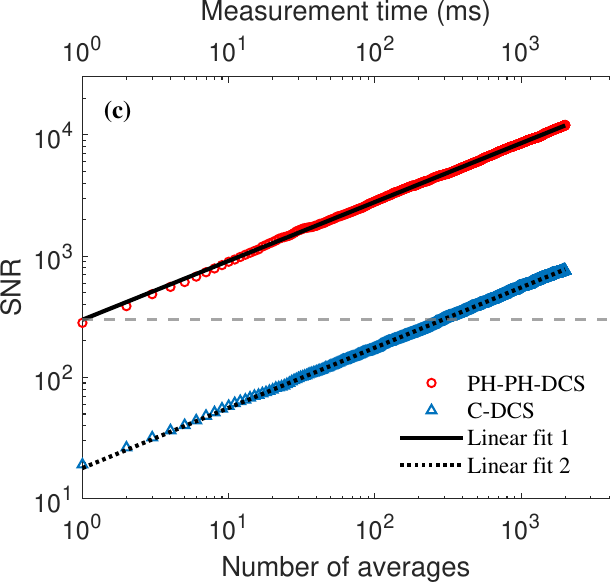}
    \end{subfigure}
    \caption{\textbf{(a) }Noise sections of interferogram (IGM) used to calculate PSDs. The main and small centerbursts and their surrounding regions containing the free induction decay are excluded. \textbf{(b)} PSDs of the noise sections of the three DCS. Three colored dashed lines represent the theoretical shot noise limits for the corresponding DCS in a single measurement (1 ms), calculated from the optical power and detector response. The shaded region from 3 to 6 MHz corresponds to the location of the interferogram spectrum. Within this region, in a single measurement, the noise power of C-PH-DCS (green) and PH-PH-DCS (red) increased by 0.5 dB and 2.5 dB, respectively, compared to C-DCS (blue). In a $\sim$2 s measurement (1999 averages), the increase was 0.75 dB and 2.4 dB, respectively. \textbf{(c)} Evolution of time-domain SNR for the PH-PH-DCS and the C-DCS with the number of averages. On the log-log scale, the linear fitting models of the two datasets are given by \textit{Fit 1} = 0.4851(4)\textit{x} + 2.475(2) and \textit{Fit 2} = 0.4962(4)\textit{x} + 1.253(1). The slopes, approximating 0.5, indicate that the SNR is proportional to the square root of the number of averages, consistent with the principle of coherent averaging. The time-domain SNR is defined as the ratio of the centerburst amplitude to the standard deviation of some segments away from the centerburst \cite{DualcombPhotoacoustic2020friedlein}.}
    \label{fig:Noise-floor}
\end{figure*}

The power spectral densities (PSDs) of the noise sections of the three DCS are shown in Fig. \ref{fig:Noise-floor}. In a single 1-ms measurement, none of them reached the shot-noise limit. This shortfall is primarily due to the optical power loss introduced by bandpass filtering. The optical power increased by a factor of about $\sim$17x from C-DCS to PH-PH-DCS, whereas the noise power only increased by a factor of 1.74x (2.4 dB). This discrepancy is likely due to quantization noise or relative intensity noise (RIN).
 
The evolution of the time-domain SNR for the PH-PH-DCS and the C-DCS as a function of the number of averages (or measurement time) is shown in Fig. \ref{fig:Noise-floor} (c). The x-axis separation of 2.463 between \textit{fit 2} and \textit{fit 1} at the same SNR level indicates that this new architecture reduces the averaging time requirement by a factor of 290, i.e., down to 1/290 of the original time. The y-axis intercepts differ by 1.222, corresponding to a 16.7-fold SNR enhancement at the same number of averages. These observations agree well with the expected performance for a multiplexing factor of \textit{m} = 20 in approximately constant-noise-limited region. 

Nevertheless, if the system operates in the shot-noise limited region \cite{UnlockingLower2023walsh}, the SNR enhancement and averaging time reduction are expected to decrease to factors of 4 and 17, respectively. More generally, for an arbitrary multiplexing factor m, the SNR scales as $\sqrt{m}$ and the averaging time reduces by a factor of \textit{m}.

\begin{figure*}[!htb]
    \captionsetup[subfigure]{labelformat=empty}
        \centering
        \begin{subfigure}{0.33\textwidth}
          \includegraphics[clip,width=\textwidth]{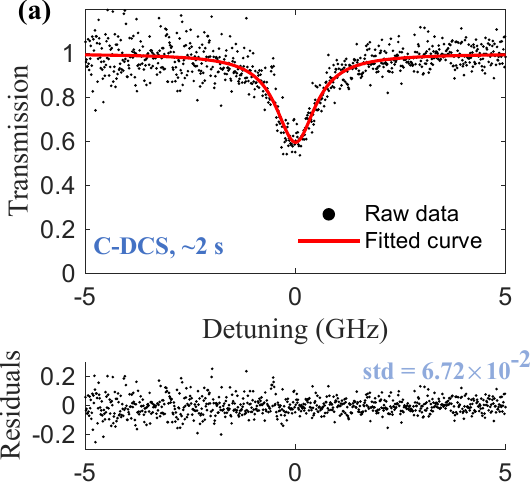}
        \end{subfigure}%
        ~~~~
        \begin{subfigure}{0.33\textwidth}
          \includegraphics[clip,width=\textwidth]{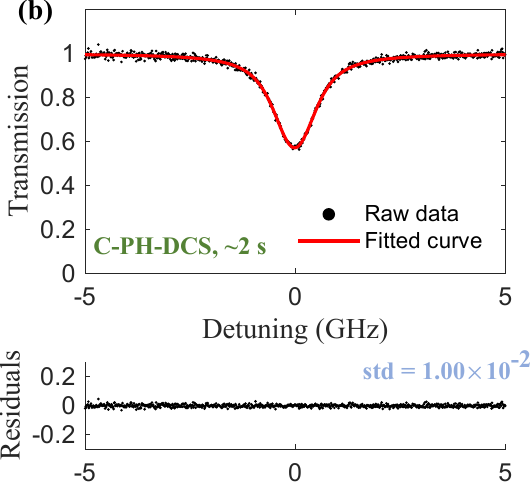}
        \end{subfigure}%
        ~~~~
        \begin{subfigure}{0.33\textwidth}
          \includegraphics[clip,width=\textwidth]{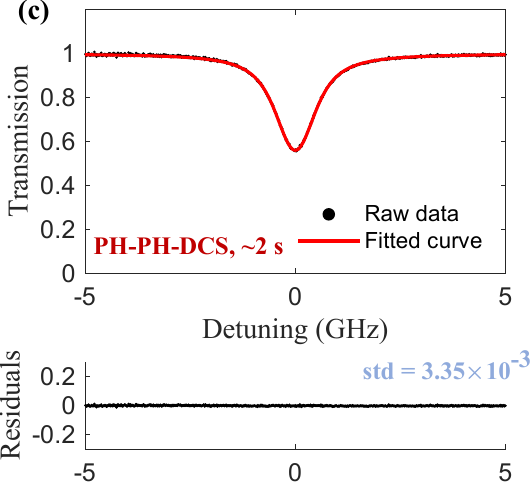}
        \end{subfigure}%
        \caption{Example $\mathrm{H^{13}C^{14}N}$ spectra relative to the center frequency of 193.5449 THz (1548.9555 nm) at a spectral resolution of 12.5 MHz, along with the corresponding weighted Voigt profile fits. \textbf{(a)} Recorded in 1.999 s (1999 averages) in the C-DCS. \textbf{(b)} Recorded in 1.999 s in the C-PH-DCS. \textbf{(c)} Recorded in 1.999 s in the PH-PH-DCS.}
    \label{fig:abpeaks of three architectures}
\end{figure*}

\subsection{Baseline Normalization and Transmission Spectrum}

As mentioned, the baselines of these spectra were normalized by the second reference channel. This is because the actual modulated phase levels deviate from the ideal values, leading to mode-to-mode variations in the spectrum, as shown in Figs. \ref{fig:IGMs&Spec of 3 archis} (d) and (f). Such variations are also commonly observed in DCS setups based on electro-optic (EO) combs \cite{NanosecondTimeresolved2024long} and certain microcombs. These high-frequency baseline fluctuations are often difficult to correct using conventional self-fitting methods \cite{IntercomparisonOpenpath2017waxman}.

Nevertheless, in PPHE-DCS, such fluctuations exhibit a periodicity of $m$-times mode spacing. Utilizing this periodic characteristic, a spectrum of PPHE-DCS can be deinterleaved into \textit{m} normal spectra, which can then be fitted and processed individually as in C-DCS, without relying on the second channel. A detailed demonstration is provided in Supplement 1. 

The enhancement in time-domain SNR is evident in the transmission spectrum. Fig. \ref{fig:abpeaks of three architectures} presents the transmission spectra acquired around the P7 absorption line of $\mathrm{H^{13}C^{14}N}$ with the three architectures, with a spectral point spacing of 12.5 MHz. Both architectures, particularly the PH-PH-DCS, demonstrate significantly lower residual levels within the same measurement time compared to the C-DCS, with the residual noise reduced by a factor of $\sim$20.

It is worth noting that the three weighted Voigt profile fits show slight differences in line depth, we attribute this to the differences in SNR and tracking bandwith of the cross-correlation function \cite{SelfCorrectionFreeRunning2025xiangze}.

\subsection{Discussion}

We expect this concept can be integrated with ultra-high-repetition-rate OFC source, such as microcombs \cite{MicroresonatorSoliton2016suh}. In that case, this PPHE-DCS can achieve a higher SNR enhancement or averaging time reduction, and become applicable to typical use cases requiring spectral sampling spacings from hundreds of MHz to the GHz range \cite{Broadband1GHz2022hoghooghi}.

Finally, it is worth discussing that the PH-OFC share an essential similarity with harmonic mode-locking: the pulse train exhibits correlation between every \textit{m}th pulse \cite{CorrelationSupermode2007gee}, whereas neighboring pulses of which are not fully correlated. Under this condition, the spectrum contains an interleaving of \textit{m} supermodes, resulting in an effective mode spacing of $f_{\mathrm{rep}}/m$ rather than $f_{\mathrm{rep}}$. These additional supermodes, apart from the dominant one, are conceptually similar to the "densified" or "self-imaging” modes here. 
Therefore, We speculate that the extra-cavity-TPM-modulating scheme, may not be the only viable approach to realize this concept. For instance, conventional harmonically mode-locked lasers aim for strong supermode suppression \cite{EchelleSpectrograph2009mcferran}. In contrast, if low supermode suppression could be achieved such that the \textit{m} supermodes have comparable intensities, it might provide a suitable source for PPHE-DCS. Moreover, in some parametrically driven Kerr soliton microresonators, two soliton states with opposite phases owing to the $Z_2$-symmetry have been observed \cite{ParametricallyDriven2024moille}, and a soliton sequence with random phase bits of $[0, 0, \pi, 0]$ has been demonstrated \cite{ParametricallyDriven2021englebert}.

\medskip
\section{Conclusion}

In summary, we demonstrate a DCS concept that enhances SNR to reduces coherent averaging time under detector dynamic range limitations by employing phase-pattern higher-repetition-rate frequency combs as sources. We reinterpret the self-imaging process of comb spectra from the perspective of mode interference among sub-pulse trains, and provide a comprehensive description of two architectures for this special form of DCS. In a proof-of-concept experiment, by 20-fold pulse multiplexing, we achieved an $\sim$16.7-fold SNR improvement or a corresponding $\sim$290-fold reduction in averaging time compared to a conventional DCS architecture, in the non-shot-noise-limited regime. These performance gains are expected to decrease to $\sim$4-fold and $\sim$17-fold, respectively, in the shot-noise-limited regime.  This approach reduces the reliance on long-term averaging, thereby enabling recovery of the inherent temporal resolution of DCS and potentially relaxing the stringent requirements on coherence time and stability of DCS systems. We expect that this concept can be combined with ultra-high-repetition-rate microcombs for deployment in applications that typically require spectral sampling spacings from hundreds of MHz to the GHz range. 

\medskip

\medskip
\noindent\textbf{Acknowledgments}

\begin{footnotesize}
\noindent This work was supported by National Key Research and Development Program of China under Grant2020YFC2200300. The authors thank Prof. Tang Li for providing the Menlo-250-WG-PM, Yatan Xiong for lending the EDFA, and Hao Li, Yinnan Chen, Zheng Liu, Yujia Ji, and Yanan Qu for their assistance with certain details.
\end{footnotesize}
\medskip

\noindent\textbf{Disclosures}

\begin{footnotesize}
\noindent The authors declare no conflicts of interest.
\end{footnotesize}
\medskip

\noindent\textbf{Data availability} 

\begin{footnotesize}
\noindent Data underlying the results presented in this paper are not publicly available at this time but may be obtained from the authors upon reasonable request.
\end{footnotesize}
\medskip

\noindent\textbf{Supplemental document}

\begin{footnotesize}
\noindent See Supplement 1 for supporting content. 
\end{footnotesize}
\medskip

\medskip
\bibliography{ref}

\end{document}


\title{Supplement 1: Reducing Averaging Time in Dual-comb Spectroscopy via Phase-Patterned Higher-Repetition-Rate Pulses}

\author{Wei Long$^{1,2,*}$, Xinru Cao$^{1,2,3}$, Xiangze Ma$^{1,2,3,4}$, Jiaqi Zhou$^{1,2,3}$, Wenbin He$^{5}$, Dijun Chen$^{1,2,3,4,*}$\\
\medskip
$^1${Wangzhijiang Innovation Center for Laser, Aerospace Laser Technology and System Department, Shanghai Institute of Optics and Fine Mechanics, Chinese Academy of Sciences, Shanghai 201800, China}\\
$^2${Shanghai Key Laboratory of All Solid-State Laser and Applied Techniques, Shanghai Institute of Optics and Fine Mechanics, Chinese Academy of Sciences, Shanghai 201800, China}\\
$^3${Center of Materials Science and Optoelectronics Engineering, University of Chinese Academy of Sciences, Beijing 100049, China}\\
$^4${Hangzhou Institute for Advanced Study, University of Chinese Academy of Sciences, Hangzhou 310024, China}\\
$^5${Russell Centre for Advanced Lightwave Science, Shanghai Institute of Optics and Fine Mechanics and Hangzhou Institute of Optics and Fine Mechanics, Chinese Academy of Sciences, Shanghai 201800, China}\\
$^{*}$longwei20@mails.ucas.ac.cn\\
$^{*}$djchen@siom.ac.cn}

\date{\today}

\maketitle

\begin{figure}[!ht]
    \centering
    \includegraphics[width=0.5\linewidth]{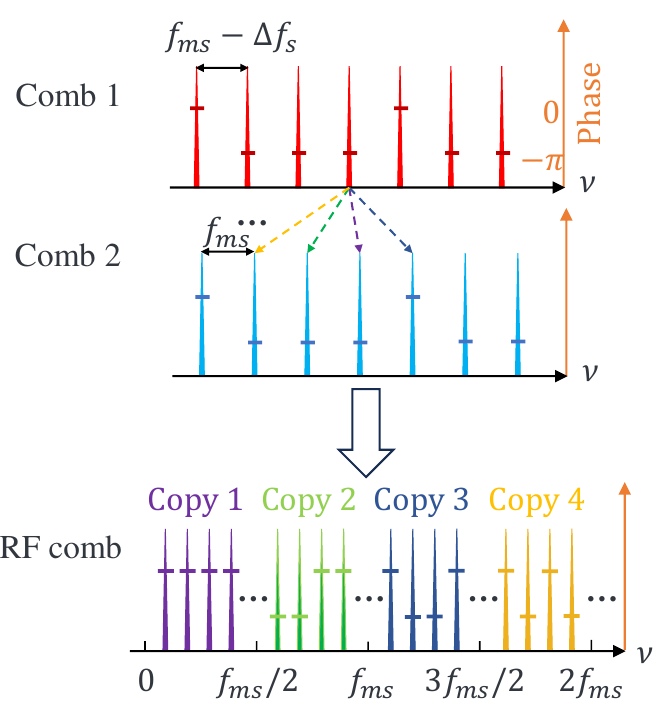}
    \caption{Illustration of spectral copies generated by the heterodyning of teeth from comb 1 with non-nearest teeth from comb 2.} 
    \label{fig:spectral copies}
\end{figure}

\begin{figure}[!ht]
    \centering
    \includegraphics[width=0.6\textwidth]{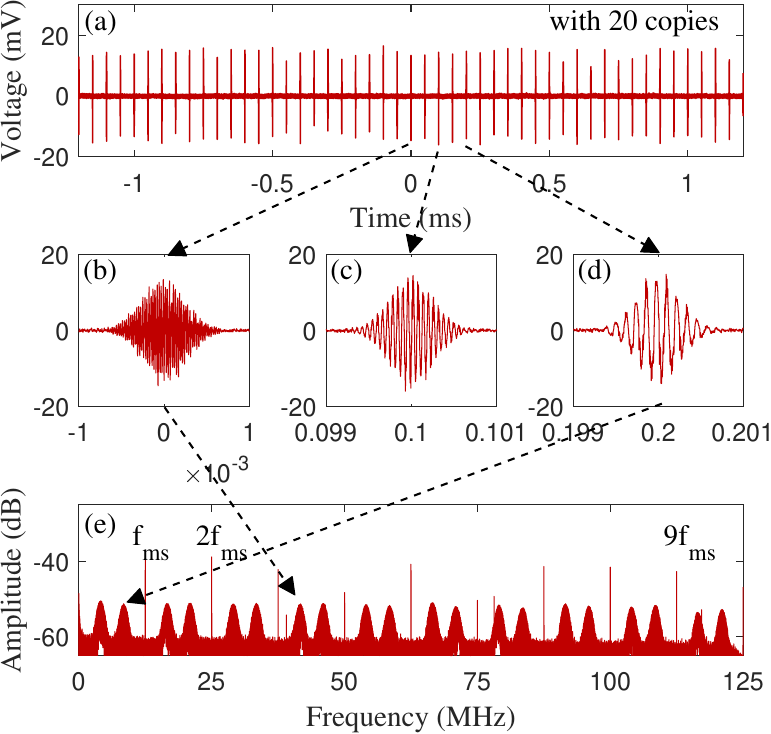}
    \caption{(a) Interferograms of the PH-PH-DCS and (e) its spectrum with 20 spectral copies. The frequencies evenly spaced in the spectrum at intervals of 12.5 MHz are harmonics of the comb mode spacing. These harmonics have negligible impact on (a) interferograms due to their low amplitude and inconsistent phase. (b–d) Centerbursts corresponding to different spectral copies, show different carrier frequencies.}
    \label{fig:20 spectral copies}
\end{figure}

The higher-order components in the interferogram of PH-PH-DCS correspond to spectral copies on the spectrum, which are illustrated in Fig. \ref{fig:spectral copies}. The different mode-phase profiles of these copies led to their self-imaging and time shift in the time domain. These properties are more evident when the two TPM signals are based on the quadratic forms of the "Talbot condition" (see Equ. 3). In that case, the phase profiles of these copies exhibit a linear dependence on the line number \textit{j}: 
\begin{equation}\label{eq:copies phase}
\begin{aligned}
\varphi_{j,h} &= (-\pi\frac{p}{m}j^2) -[- \pi\frac{p}{m}(j-h)^2]\\
&= -2\pi\frac{p}{m}hj+\pi\frac{p}{m}h^2,
\end{aligned}
\end{equation}
where \textit{h} denotes the detuning line number and belongs to the range $[-m/2,m/2)$. The first term, $-2\pi\frac{p}{m}hj$, mirrors the time-shift property of the Fourier transform, $\mathcal{F}\left\{f(t - t_0)\right\}= e^{-i2\pi jf_{\mathrm{ms}}t_0}F(\omega)$, indicating a time shift $t_h=h\cdot\frac{p}{mf_{\mathrm{ms}}}$ for these copies. The second term, $\pi\frac{p}{m}h^2$, signifies a carrier-envelope-phase (CEP) shift that varies with \textit{h}, which does not impact the centerburst distribution. In a basic scenario with \textit{p} = 1, the centerbursts corresponding to each copy are uniformly distributed across the interferogram within a period, spaced at intervals of $1/(mf_{\mathrm{ms}})$ . This conclusion holds true for any \textit{p} setting, given that \textit{p} and \textit{m} are coprime and possess opposite parity, indicating that the interference electric fields associated with multiple copies are dispersed into \textit{m} parts in the time domain, rather than being concentrated into a single centerburst as in C-DCS. Figure \ref{fig:20 spectral copies}(a) shows the interferogram obtained during the PH-PH-DCS experiment when using a 125 MHz electronic low-pass filter, resulting in the inclusion of 20 spectral copies. The centerburst repetition rate is 20 kHz rather than 1 kHz, and each centerburst has a different carrier frequency. 

These higher-order components can be further utilized to perform spectral-copy averaging to improve SNR, as demonstrated by M. Walsh et al \cite{UnlockingLower2023walsh}. The averaging method can be replaced by simply undersampling at a sampling frequency of $f_{ms}/2$, in which case the characteristics of the resulting interferogram are expected to be the same as those in C-PH-DCS.

\begin{figure}[!ht]
    \centering
    \includegraphics[width=0.9\linewidth]{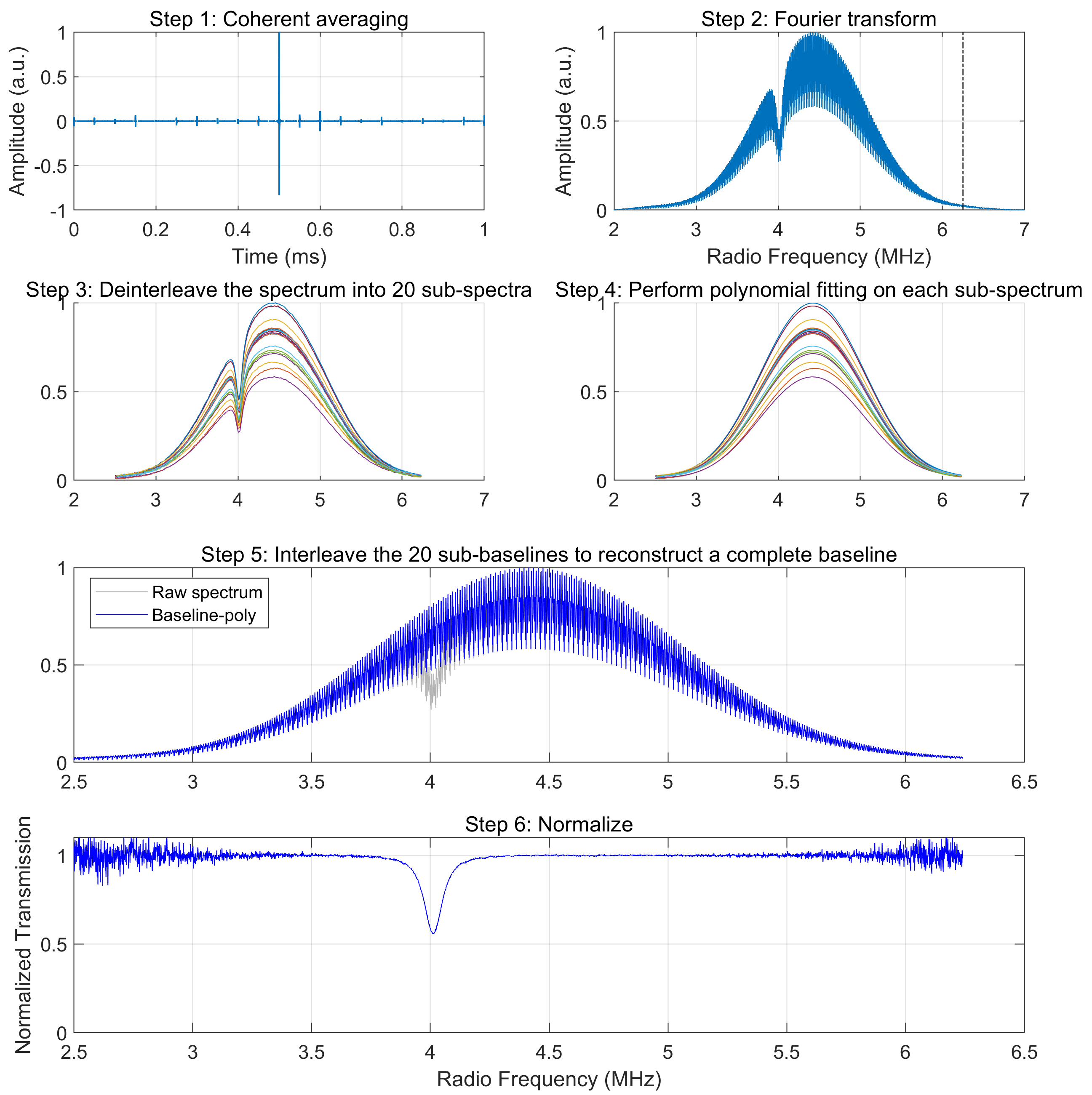}
    \caption{Demonstration of spectral baseline normalization.} 
    \label{fig:baseline_nor}
\end{figure}
A baseline-normalized method that not relying on the second reference channel is shown in Fig. \ref{fig:baseline_nor}. This method takes advantage of the periodicity of the mode-to-mode-amplitude variation in this technique. Due to this periodicity, the spectrum can be regarded as interlaced by 20 normal spectra with a mode spacing of 20 kHz (in the RF domain, and is of 250 MHz in the optical frequency domain). Therefore, the spectrum can be decomposed into 20 sub-spectra, and perform polynomial fitting on each sub-spectrum as that in conventional DCS normalization to obtain 20 sub-baselines. In this demonstration, the poly degree is set to 10 and 37 points around the absorption peak were excluded in polynomial fitting. 
Then,  a complete baseline for normalization can be reconstructed by interleaving these sub-baselines. The resulting normalized transmission spectrum shows good agreement with that obtained using the reference channel, with a standard deviation of the difference being $3.11\times10^{-3}$, which is close to the residual additive noise level of being $3.35\times10^{-3}$.

\smallskip
\smallskip
\bibliography{SIref}